\documentclass[10pt]{article}
\usepackage{latexsym,graphicx}
\usepackage{latexsym,graphicx}
\newcommand{\be}{\begin{equation}}
\newcommand{\ee}{\end{equation}}
\def\n{\noindent}
\catcode `\@=11 \catcode `\@=12
\begin{document}
\begin{center}
\large{\bf {Bianchi Type III Anisotropic Dark Energy Models with Constant Deceleration Parameter}} \\
\vspace{10mm}
\normalsize{Anil Kumar Yadav\footnote{corresponding author} and Lallan Yadav$^2$}\\
\vspace{4mm}
\normalsize{$^1$Department of Physics, Anand Engineering
College,Keetham, Agra-282 007, India} \\
\normalsize{E-mail: abanilyadav@yahoo.co.in, akyadav@imsc.res.in}\\
\vspace{2mm}
\normalsize{$^2$ Department of Physics, D. D. U. Gorakhpur University, Gorakhpur - 273 009}
\normalsize{E-mail: nisaly06@rediffmail.com}
%\vspace{5mm}
%\normalsize{}
%\vspace{5mm}
%\normalsize{}
\end{center}
\vspace{10mm}
%\date{}
%\maketitle
%\newpage
\begin{abstract} 
\n The Bianchi type III dark energy models with constant deceleration parameter is investigated.
The equation of state parameter $\omega$ is found to be time dependent and its existing range 
for this model is consistent with the recent observations of SN Ia data, SN Ia data (with CMBR anisotropy) 
and galaxy clustering statistics. The physical aspects of the dark energy models is discussed.   
\end{abstract}
\smallskip
\n Keywords : Bianchi type III Universe, Dark Energy, EoS Parameter\\ 
%%%%%%%%%%%%%%%%%%%%%%%%%%%%%%%%%%%%%%%%%%%%%%%%%%%%%%%%%%%%%%%%%%%%%%%%%%%%%%%%%%%
%%%%%%%%%%%%%%%%%%%%%%%%%%%%%%%%% section 1 Introduction %%%%%%%%%%%%%%%%%%%%%%%%%%
\section{Introduction}
Arguments have been put forward that we live in a spatially flat low matter density universe which is currently undergoing a period of accelerating expansion. If the observational evidence upon which these claims are based are rainforced and strengthened by future experiments, the implications for cosmology will be incredible. It could then appear that the cosmological fluid is dominated by some sort of fantastic energy density, which has negative pressure and has just begun to play significant role today. 
Recent years have witnessed the emergence of the idea of an accelerating universe. Therefore, due to 
some observational results \cite{ref1}$-$\cite{ref4} it is now established that universe is accelerating. 
This signifies a remarkable shift in cosmological research from expanding universe to accelerated expanding 
universe. Now, the problem lies in detecting an exotic type of unknown repulsive force, termed as dark energy.
The detection of dark energy would be a new clue to an old puzzle: the gravitational effect of the zero - 
point energies of particles and fields. The total with other energies, that are close to homogeneous 
and nearly independent of time, acts as dark energy. The paramount characteristic of the dark energy is a 
constant or slightly changing energy density as the universe expands, but we do not know the nature of dark 
energy very well \cite{ref5}$-$\cite{ref13}. Dark Energy (DE) has been conventionally 
characterized by the equation of state (EoS) parameter $\omega=p/\rho$ which is not necessarily constant.
The simplest DE candidate is the vacuum energy $(\omega = -1)$, which is mathematically equivalent to the 
cosmological constant $(\Lambda)$. The other conventional alternatives, which can be described by minimally 
coupled scalar fields, are quintessence $(\omega>-1)$, phantom energy $(\omega<-1)$ and quintom 
(that can across from phantom region to quintessence region) as evolved and have time dependent EoS parameter.
Some other limits obtained from observational results coming from SN Ia data \cite{ref14} and SN Ia data 
collaborated with CMBR anisotropy and galaxy clustering statistics \cite{ref15} are $-1.67<\omega<-0.62$ and 
$-1.33<\omega<-0.79$ respectively. However, it is not at all obligatory to use a constant value of $\omega$.
Due to lack of observational evidence in making a distinction between constant and variable $\omega$, usually 
the equation of state parameter is considered as a constant \cite{ref16,ref17} with phase wise value $-1$, $0$, 
$-1/3$ and $+1$ for vacuum fluid, dust fluid, radiation and stiff fluid dominated universe, respectively.  
But in general, $\omega$ is a function of time or redshift \cite{ref18}$-$\cite{ref20}. For instance, quintessence 
models involving scalar fields give rise to time dependent EoS parameter $\omega$ \cite{ref21}$-$\cite{ref24}. 
Also some literature is available on models with varying fields, such as cosmological model with variable equation 
of state parameter in Kaluza-Klein metric and wormholes \cite{ref25,ref26}.
In recent years various form of time dependent $\omega$ have been used for variable $\Lambda$ models \cite{ref27,ref28}.
Recently Ray et al \cite{ref29}, Akarsu and Kilinc \cite{ref30} and Yadav et al \cite{ref31} have studied  dark energy model with variable EoS parameter.
\par 
Spatially homogeneous and anisotropic cosmological models play a significant role in description of the large
 scale behaviour of universe. Bianchi type III cosmological model in presence of dark energy have been studied 
in general relativity by numerous authors. Lorentz \cite{ref32} has presented a model with dust and cosmological 
constant. Chakraborty and Chakraborty have given a bulk viscous cosmological model with variable 
$G$ and $\Lambda$ \cite{ref33}. Singh et al \cite{ref34} have investigated a model with variable 
$G$ and $\Lambda$ in presence of perfect fluid by assuming a conservation law of energy-momentum tensor. 
Recently, Tiwari \cite{ref35} has studied a model in presence of perfect fluid and time dependent $\Lambda$
with constant deceleration parameter. Bali and Tinkar \cite{ref36} have investigated a model in the 
presence of bulk viscous borotropic fluid with variable $G$ and $\Lambda$. Unlike Robertson-Walkar metric, 
Bianchi type III can admit a dark energy that yields an anisotropic EoS parameter according to their 
characteristics. The cosmological data from the large scale structure \cite{ref37} and type Ia 
supernova \cite{ref3,ref4} observations-do not rule out the possibility of anisotropic dark energy 
either \cite{ref38,ref39}.

In this paper, we have investigated the anisotropic DE 
models with variable $\omega$. This paper is organized as follows: 
In section 2, the metric and field equations are described. The solution of field equations are presented in section 3 and section 4 concludes the findings.

%%%%%%%%%%%%%%%%%%%%%%%%%%%%%%%%%%%%%%%%%%%%%%%%%%%%%%%%%%%%%%%%%%%%%%%%%%%%%%%%%%%
%%%%%%%%%%%%%%%%%%%%%%%%%%%%%%%  SECTION 2  %%%%%%%%%%%%%%%%%%%%%%%%%%%%%%%%%%%%%%%%
\section{The Metric and Field  Equations}
We consider Bianchi type III metric in the form
\begin{equation}
\label{eq1}
ds^2 = -dt^2 +A^{2}dx^2 +B^{2} e^{2\alpha x}dy^2 + C^2dz^2
\end{equation}
where A, B and C are the function of t only.\\
The simplest generalisation of EoS parameter of perfect fluid may be to determine 
the EoS parameter separately on each spatial axis by preserving the diagonal form 
of the energy-momentum tensor in a consistent way with the considered metric. 
Thus, the energy momentum tensor of fluid is taken as
\begin{equation}
\label{eq2}
T_{i}^{j} = diag\left[T_{0}^{0}, T_{1}^{1}, T_{2}^{2}, T_{3}^{3}\right] 
\end{equation}
Then, one may parametrize it as follows,
\[
 T_{i}^{j} = diag\left[\rho, -p_{x}, -p_{y}, -p_{z}\right] 
= diag\left[1, -\omega_{x}, -\omega_{y}, -\omega_{z}\right]\rho 
\]
\begin{equation}
\label{eq3}
= diag\left[1, -\omega, -\left(\omega + \delta \right), -\left(\omega + \gamma\right)\right]\rho   
\end{equation}
where $ \rho $ is the energy density of fluid,; $p_{x}$, $p_{y}$ and $p_{z}$ are the pressures and $\omega_{x}$, 
$\omega_{y}$ and $\omega_{z}$ are the directional EoS parameters along the x, y and z axes respectively.
$\omega$ is the derivation-free EoS parameter of the fluid. We have parametrized the deviation from isotropy
by setting $\omega_{x} = \omega$ and then introducing skewness parameter $\delta$ and $\gamma$ that are deviation from $\omega$ along y and z axis respectively.\\
The Einstein field equations, in gravitational units ($8\pi G =1$ and $c=1$), are
\begin{equation}
\label{eq4}
R_{ij} - \frac{1}{2}Rg_{ij} = -T_{ij} 
\end{equation}
where the symbols have their usual meaning.\\
In a comoving co-ordinate system, Einstein's field equation (\ref{eq4}), for the anisotropic
 Bianchi type III metric (\ref{eq1}), 
in case of (\ref{eq3}), lead to the following system of equations
\begin{equation}
\label{eq5}
\frac{A_{4}B_{4}}{AB}+\frac{A_{4}C_{4}}{AC}+\frac{B_{4}C_{4}}{BC}-\frac{\alpha^2}{A^2}=\rho
\end{equation}
\begin{equation}
\label{eq6}
\frac{B_{44}}{B}+\frac{C_{44}}{C}+\frac{B_{4}C_{4}}{BC}=-\omega\rho
\end{equation}
\begin{equation}
\label{eq7}
\frac{A_{44}}{A} + \frac{C_{44}}{C}+\frac{A_{4}C_{4}}{AC} = -\left(\omega + \delta\right)\rho 
\end{equation}
\begin{equation}
\label{eq8}
\frac{A_{44}}{A}+\frac{B_{44}}{B}+\frac{A_{4}B_{4}}{AB}-\frac{\alpha^2}{A^2}=-\left(\omega + \gamma\right)\rho 
\end{equation}
\begin{equation}
\label{eq9}
\alpha\left(\frac{A_{4}}{A}-\frac{B_{4}}{B}\right)=0 
\end{equation}
Here, the sub indices 4 in A, B, C and elsewhere denote differentiation with respect to t.\\
Integrating equation (\ref{eq9}), we obtain
\begin{equation}
\label{eq10}
A = kB
\end{equation}
where $k$ is the positive constant of integration. We substitute the value of equation (\ref{eq10}) 
in equation (\ref{eq7}) and subtract the result from equation (\ref{eq6}), we obtain that the skewness 
parameter on y-axis is null i.e.
$$\delta = 0$$
Thus system of equations from (\ref{eq5}) - (\ref{eq9}) may be reduce to
\begin{equation}
\label{eq11}
\frac{A_{4}^2}{A^2}+2\frac{A_{4}C_{4}}{AC}-\frac{\alpha^2}{A^2}=\rho
\end{equation}
\begin{equation}
\label{eq12}
\frac{A_{44}}{A}+\frac{C_{44}}{C}+\frac{A_{4}C_{4}}{AC}=-\omega\rho
\end{equation}
\begin{equation}
\label{eq13}
2\frac{A_{44}}{A}+\frac{A_{4}^2}{A^2}-\frac{\alpha^2}{A^2}=-\left(\omega+\gamma\right)\rho
\end{equation}
Now we have three linearly independent equations (\ref{eq11}) - (\ref{eq13}) and five unknown 
parameters $\left(A, B, \omega, \rho, \gamma \right) $.
Two additional constraint relating these parameters are required to obtain explicit solutions of the system.\\

\n (i) The law of variation of Hubble's parameter that yields a constant 
value of deceleration parameter. Such type of relation have already been considered 
by Berman \cite{ref40} for solving FRW models. Later on many authors 
(Singh et al \cite{ref41,ref42} and references therein) 
have studied flat FRW and Bianchi type models by using the special law of Hubble parameter that yields 
constant value of deceleration parameter.\\

\n (ii) We assume that the expansion $(\theta)$ in the model is proportional to the shear ($\sigma$). This condition leads to
\begin{equation}
\label{eq14}
B=C^n
\end{equation}
where n is proportionality constant. The motive behind assuming condition (ii) is explained with reference to Throne $(1967)$ \cite{ref43}, the observations of the velocity red-shift relation for extragalatic sources suggest that Hubble expansion of the universe is isotropy today within $\approx 30$ percent (Kantowski and Sachs \cite{ref44}; Kristian and Sachs \cite{ref45}). To put more precisely, red-shift studies place the limit\\
$$\frac{\sigma}{H}\leq0.3$$
\n on the ratio of shear $\sigma$ to Hubble constant H in the neighbourhood of our galaxy today. Collin et al \cite{ref46} have pointed out that for spatially homgeneous metric, the normal congruence to the homogeneous expansion satisfies that the condition $\frac{\sigma}{\theta}$ is contant.\\

\n The average scale factor of Bianchi type III metric is given by
\begin{equation}
\label{eq15}
R = \left(ABCe^{\alpha x}\right)^\frac{1}{3} 
\end{equation}
We define, the generalised mean Hubble's parameter $H$ as
\begin{equation}
\label{16}
H = \frac{1}{3}\left(H_{1} + H_{2} + H_{3}\right) 
\end{equation}
where $H_{1} = \frac{A_{4}}{A}$, $H_{2} = \frac{B_{4}}{B}$ and $H_{3} = \frac{C_{4}}{C}$ are the directional Hubble's 
parameter in the direction of x, y and z respectively.\\
Equation (\ref{eq15}), may be reduces to
\begin{equation}
\label{eq17}
H = \frac{R_{4}}{R} = \frac{1}{3}\left(\frac{A_{4}}{A} +\frac{B_{4}}{B}+\frac{C_{4}}{C}\right) 
\end{equation}
Since the line-element (\ref{eq1}) is completely characterized by Hubble's parameter H.
Therefore, let us consider that mean Hubble parameter H is related to average scale factor $R$ 
by following relation
\begin{equation}
\label{eq18}
H = k_{1}R^{-s} 
\end{equation}
where $k_{1} > 0$ and $s \geq 0$, are constant. \\
The deceleration parameter is defined as
\begin{equation}
\label{eq19}
q = -\frac{RR_{44}}{R_{4}^2} 
\end{equation}
From equations (\ref{eq17}) and (\ref{eq18}), we get
\begin{equation}
\label{eq20}
R_{4} = k_{1}R^{-s+1} 
\end{equation}
\begin{equation}
\label{eq21}
R_{44} = -k_{1}^{2}(s-1)R^{-2s+1} 
\end{equation}
Using equations (\ref{eq15}) and (\ref{eq19}), equation (\ref{eq17}) leads to
\begin{equation}
\label{eq22}
q = s-1~~(constant) 
\end{equation}
The sign of $q$ indicates whether the model inflates or not. The positive sign of $q$ corresponds to
standard decelerating model where as the negative sign of $q$ indicates inflation. However the current 
observations of SN Ia and CMBR favour accelerating models i.e. $q < 0$.\\
From equation (\ref{eq19}), we obtain the law of average scale factor R as
\begin{equation}
\label{eq23}
R =  \left[ \begin{array}{ll}
            \left(Dt + c_{1}\right)^\frac{1}{s}  & \mbox { when $s \neq 0$}\\
 c_{2}e^{k_{1}t}                                      & \mbox { when $s = 0$} 
            \end{array} \right.
\end{equation}
where $c_{1}$ and $c_{2}$ are the constant of integration.\\
From equation (\ref{eq22}), for $s \neq 0$, it is clear that the condition for expansion 
of universe is $ s > 0$ i.e. $ q+1>0 $. Therefore for expansion model of universe the deceleration 
parameter (q) should be greater than $-1$.\\
%%%%%%%%%%%%%%%%%%%%%%%%%%%%%%%%%%%%%%%%%%%%%%%%%%%%%%%%%%%%%%%%%%%%%%%%%%%%%%%%%%%
%%%%%%%%%%%%%%%%%%%%%%%%%%%%%%%  SECTION 3  %%%%%%%%%%%%%%%%%%%%%%%%%%%%%%%%%%%%%%%%
\section{Solutions of the Field  Equations and Discussion}
\subsection{Case(i): when $s\neq0$}
Equations (\ref{eq9}), (\ref{eq17}) and (\ref{eq23}) lead to
\begin{equation}
\label{eq24}
B = l\left(Dt + c_{1}\right)^{\frac{1}{r}} 
\end{equation}
Equation (\ref{eq14}) and (\ref{eq24}) lead to
\begin{equation}
\label{eq25}
C=l_{1}\left(Dt+c_{1}\right)^\frac{1}{rn} 
\end{equation}
From equations (\ref{eq10}) and (\ref{eq24}), we obtain
\begin{equation}
\label{eq26}
A = L\left(Dt + c_{1}\right)^{\frac{1}{r}}  
\end{equation}
where $c_{o}$ is the constant of integration and $l=(c_{0})^\frac{3n}{2n+1}$,~ $ L = kl $,~ $l_{1}=l^\frac{1}{n}$,~ $r=\frac{(2n+1)s}{3n}$.\\
Thus the Hubble's parameter $H$, scalar of expansion $\theta$ and shear scalar $\sigma$ are given by
\begin{equation}
\label{eq27}
H = \frac{k_{1}}{(Dt+c_{1})} 
\end{equation}
\begin{equation}
\label{eq28}
\theta = 3H = \frac{3k_{1}}{(Dt+c_{1})} 
\end{equation} 
\begin{equation}
\label{eq29}
\sigma^2=\frac{D^{2}(n-1)^2}{3n^2r^2 (Dt+c_{1})^2}
\end{equation}
Equations (\ref{eq28}) and (\ref{eq29}) lead to
\begin{equation}
\label{30}
\frac{\sigma}{\theta}=\frac{D(n-1)}{\sqrt{3}nrk_{1}} 
\end{equation}
Using equations (\ref{eq11}), (\ref{eq25}) and (\ref{eq26}), the energy density of the 
fluid is obtained as
\begin{equation}
\label{eq31}
\rho=\frac{D^{2}(n+2)}{nr^{2}\left(Dt+c_{1}\right)^2}-\frac{\alpha^2}{L^{2}\left(Dt+c_{1}
\right)^\frac{2}{r}}
\end{equation}
Using equations (\ref{eq12}), (\ref{eq25}), (\ref{eq26}) and (\ref{eq31}), the equation of state 
parameter $\omega$ is obtained as
\begin{equation}
\label{eq32}
\omega = \frac{\frac{D^{2}\left[n^2+(n+1)(1-rn)\right]}{n^2r^2\left(Dt+c_{1}\right)^2}}
{\frac{\alpha^2}{L^2\left(Dt+c_{1}\right)^\frac{2}{r}}-\frac{D^2(n+2)}{nr^2\left(Dt+c_{1}\right)^2}}
\end{equation}
Using equations (\ref{eq13}), (\ref{eq26}), (\ref{eq31}) and (\ref{eq32}), the skewness parameter $\gamma$
(i.e. deviation from $\omega$ along z-axis) is given by
\begin{equation}
\label{eq33}
\gamma=\frac{\frac{D^2\left[n^2(2-r)+n(r-1)-1\right]}{n^2r^2\left(Dt+c_{1}\right)^2}-\frac{\alpha^2}{L^2
\left(Dt+c_{1}\right)^\frac{2}{r}}}{\frac{\alpha^2}{L^2\left(Dt+c_{1}\right)^\frac{2}{r}}-
\frac{D^2(n+2)}{nr^2\left(Dt+c_{1}\right)^2}}
\end{equation}
\begin{figure}
\begin{center}
\includegraphics[width=4.0in]{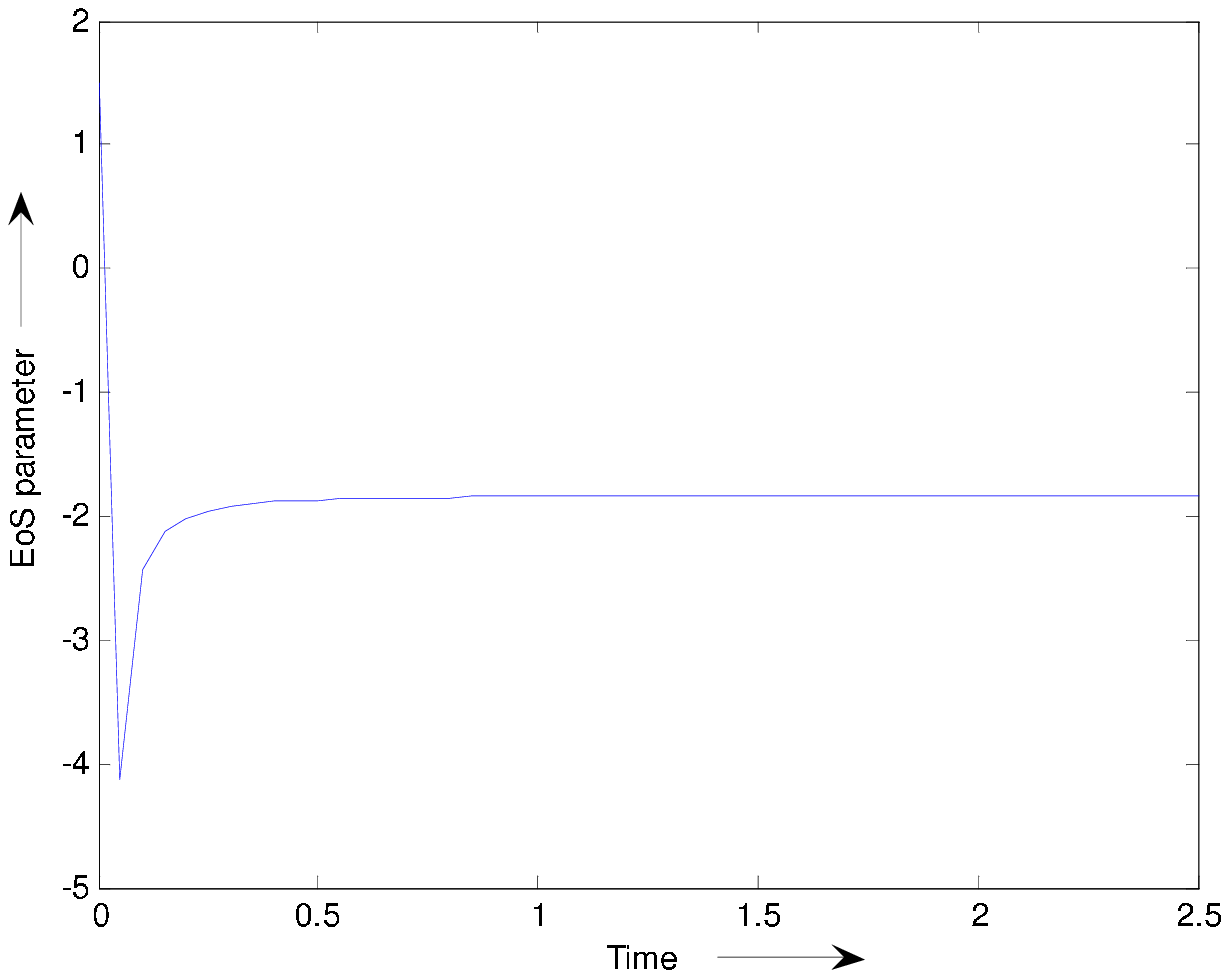} 
\caption{The plot of EoS parameter ($\omega$) vs. time (t) with $D=1$,~ $n=0.25$, $s=0.5$~$C_{1}=.05$,~
$\alpha=0.3$  and $L=1.5$}
\label{fg:anil28fig1.eps}
\end{center}
\end{figure}
From equation (\ref{eq32}), it is observed that the equation of state parameter $\omega$ is time dependent, 
it can be function of redshift z or scale factor R as well. The redshift dependence of $\omega$ can be linear
like $\omega(z)=\omega_{0}+\omega^{1}z$ with $\omega^{1} = \left(\frac{d\omega}{dz}\right)_{z}=0$ \cite{ref47,ref48} or 
nonlinear as $\omega(z)=\omega_{0}+\frac{\omega_{1}z}{1+z}$ \cite{ref49,ref50}. 
The SN Ia data suggests that $-1.67<\omega<-0.62$ \cite{ref14} while the limit imposed on $\omega$ by a combination 
of SN Ia data (with CMB anisotropy) and galaxy clustering statistics is $-1.33<\omega<-0.79$ \cite{ref15}. 
So, if the present work is compared with experimental results mentioned above then, one can conclude that 
the limit of $\omega$ provided by equation (\ref{eq32}) may accommodated with the acceptable range of EoS parameter. 
Also we see that for $s=\frac{3(n^2+n+1)}{(n+1)(2n+1)}$, the $\omega$ vanishes. Thus for 
this particular value of s, our model represents dusty universe.\\
\par 
The variation of equation of state parameter $(\omega)$ with cosmic time (t) is clearly shown 
in Fig. 1, as a representative case with appropriate choice 
of constants of integration and other physical parameters using reasonably well known situations. From Fig. 1, we conclude that in early stage, the EoS parameter $\omega$ was positive (i.e, the universe was matter dominated) and at 
late time it is evolving with negative value (i.e. at the present time). The earlier real matter later on converted to the dark energy dominated phase of universe.\\ 

In absence of any curvature, matter energy density ($\Omega_{m}$) and dark energy density ($\Omega_{\Lambda}$)  are related by the equation
\begin{equation}
\label{eq34}
\Omega_{m}+\Omega_{\Lambda}=1
\end{equation}
where $\Omega_{m}=\frac{\rho}{3H^2}$ and $\Omega_{\Lambda}=\frac{\Lambda}{3H^2}$\\
Thus equation (\ref{eq34}), reduce to
\begin{equation}
\label{eq35}
\frac{\rho}{3H^2}+\frac{\Lambda}{3H^2}=1  
\end{equation}
Using equations (\ref{eq27}) and (\ref{eq31}), in equation (\ref{eq35}), the cosmological constant is obtained as
\begin{equation}
\label{eq36}
\Lambda=\frac{3k_{1}^2}{(Dt+c_{1})^2}-\frac{D^2(n+2)}{nr^2(Dt+c_{1})^2}+\frac{\alpha^2}{L^2(Dt+c_{1})^\frac{2}{r}}
\end{equation}
Measurements of redshift of Ia supernova SN 1997 ff indicate the present acceleration of universe. 
Recently, Carmeli and Kuzmenko \cite{ref51} have shown that the cosmological relativistic theory predicts the value for cosmological constant $\Lambda=1.934\times10^{-35}s^{-2}$. This value of $\Lambda$ is excellent in agreement with the measurements recently obtained by the High-z Supernovae Cosmology Project. 
Recent supernovae Ia \cite{ref3,ref4} observations suggest that the cosmological constant is decreasing function of time and they approach to small positive values as time increases (i.e. the present epoch).
From equation (\ref{eq36}), we observe that cosmological constant ($\Lambda$) decreases as time increases.
\subsection{Case(ii): when $s=0$}
Equations (\ref{eq9}), (\ref{eq14}), (\ref{eq17}) and (\ref{eq23}) lead to
\begin{equation}
\label{eq37}
B=l_{0}e^{k_{2}t}
\end{equation}
Equations (\ref{eq14}) and (\ref{eq37}) lead to
\begin{equation}
\label{eq38}
C=l_{2}e^{\frac{k_{2}}{n}t}
\end{equation}
From equations (\ref{eq10}) and (\ref{eq37}), we obtain
\begin{equation}
\label{eq39}
A=L_{0}e^{k_{2}t} 
\end{equation}
where $\l_{0}$ is constant of integration and $L_{0}=kl_{0}$,~ $l_{2}=l_{0}^\frac{1}{n}$.,~
$k_{2}=\frac{3nk_{1}}{2n+1}$.\\
Thus the Hubble's parameter $H$, scalar of expansion $\theta$ and shear scalar $\sigma$ are given by
\begin{equation}
\label{eq40}
H = \frac{(2n+1)k_{2}}{3n}
\end{equation}
\begin{equation}
\label{eq41}
\theta = \frac{(2n+1)k_{2}}{n}
\end{equation}
\begin{equation}
\label{eq42}
\sigma^2=\frac{k_{2}^2(n-1)^2}{3n^2}
\end{equation}
Equations (\ref{eq41}) and (\ref{eq42}) lead to
\begin{equation}
\label{eq43}
\frac{\sigma}{\theta}=\frac{(n-1)}{\sqrt{3}(2n+1)}
\end{equation}
Using equations (\ref{eq11}), (\ref{eq38}) and (\ref{eq39}), the energy density of the 
fluid is obtained as
\begin{equation}
\label{eq44}
\rho=\frac{(n+2)k_{2}^2L_{0}^2-n\alpha^{2}e^{-2k_{2}t}}{nL_{0}^2} 
\end{equation}
Using equations (\ref{eq12}), (\ref{eq38}), (\ref{eq39}) and (\ref{eq44}), the equation of state 
parameter $\omega$ is obtained as
\begin{equation}
\label{eq45}
\omega=\frac{\left(n^2+n+1\right)L_{0}^2k_{2}^2}{n\left[n\alpha^2e^{-2k_{2}t}-(n+2)k_{2}^2L_{0}^2\right]}
\end{equation}
\begin{figure}
\begin{center}
\includegraphics[width=4.0in]{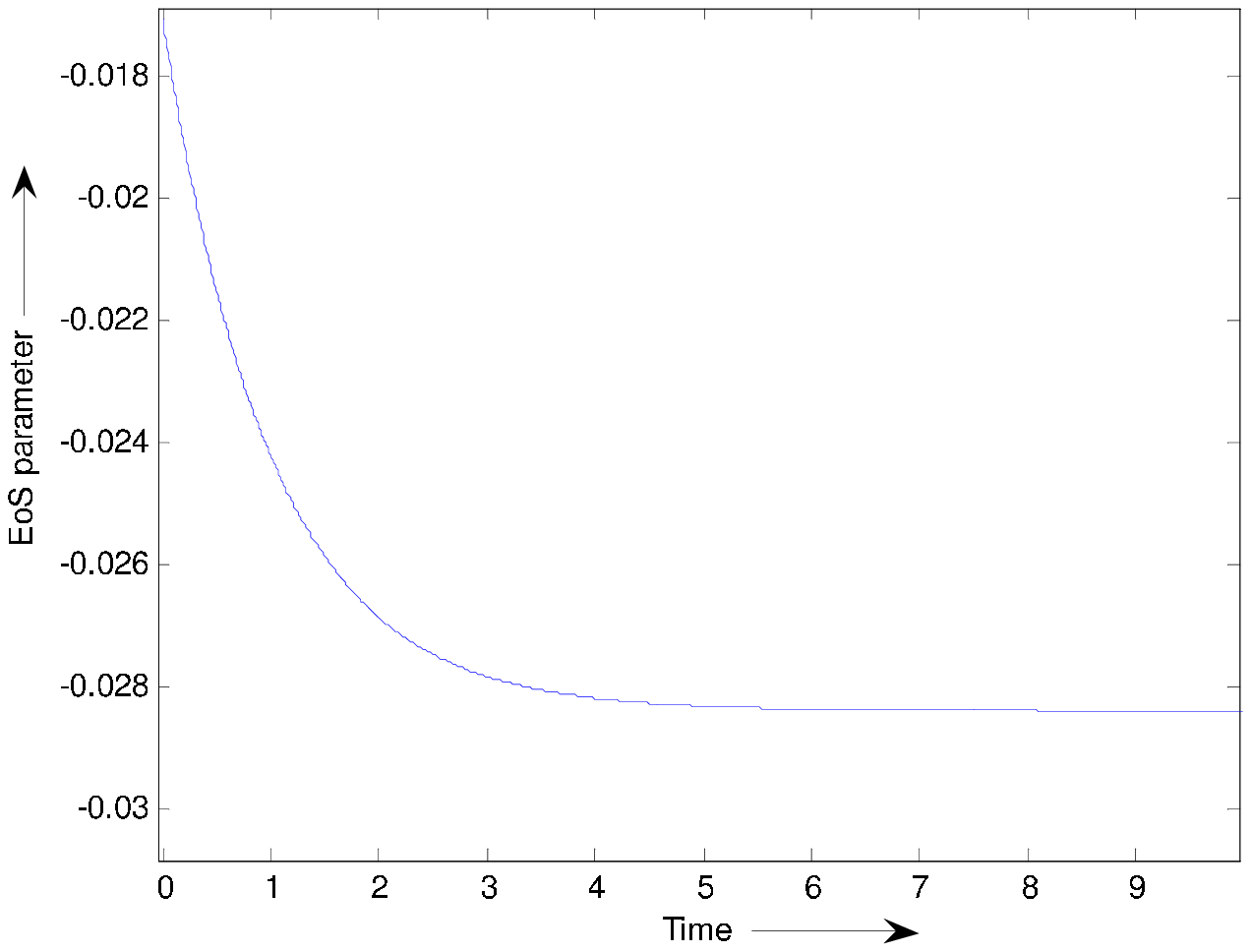} 
\caption{The plot of EoS parameter ($\omega$) vs. time (t) with $\alpha=0.3$,~ $n=0.25$,~$k_{2}=0.5$ and $L_{0}=0.10$}
\label{fg:anil28fig2.eps}
\end{center}
\end{figure} 
Using equations (\ref{eq13}), (\ref{eq39}), (\ref{eq44}) and (\ref{eq45}), the skewness parameter $\gamma$ (i.e. deviation from $\omega$ along z-axis) is given by
\begin{equation}
\label{eq46}
\gamma=\frac{(2n^2-n-1)L_{0}^2k_{2}^2-n^2\alpha^2e^{-2k_{2}t}}{n\left[n\alpha^2e^{-2k_{2}t}-
(n+2)k_{2}^2L_{0^2}\right]} 
\end{equation}
The Eos parameter $\omega$ is found to be negative and its time varying nature is clearly shown in Fig. 2.\\
Using equations (\ref{eq40}) and (\ref{eq44}), in equation (\ref{eq35}), the cosmological constant is obtained as
\begin{equation}
\label{eq47}
\Lambda=\frac{3(2n+1)^2K_{2}}{9n^2}-\frac{(n+2)k_{2}L_{0}^2-n\alpha^2e^{-2k_{2}t}}{nL_{0}^2}
\end{equation}
From equation (\ref{eq47}), it is clear that cosmological constant $\Lambda$ is decreasing function of time and approaches to small positive value at late time which is supported by results from supernova observations recently obtained by High-z Supernova Team and Supernova cosmological project \cite{ref3,ref4}. 
%%%%%%%%%%%%%%%%%%%%%%%%%%%%%%%%%%%%%%%%%%%%%%%%%%%%%%%%%%%%%%%%%%%%%%%
%%%%%%%%%%%%%%%%%%%%%%%%%%  SECTION 4  %%%%%%%%%%%%%%%%%%%%%%%%%%%%%%%%%%
\section{Concluding Remarks}
In this paper, we have studied anisotropic DE models with variable EoS parameter $\omega$, considering two 
cases, $(3.1)$ and $(3.2)$ for $s\neq0$ and $s=0$ respectively. It is observed that in both 
cases, EoS parameter $\omega$ is variable function of time which has been supported by recent observations 
\cite{ref14,ref15} and work of several authors (for example, see Refs. \cite{ref21}$-$\cite{ref31}).
In case (i), we have shown that, in early stage, the equation of state parameter $\omega$ was positive 
i.e. the universe was matter dominated and at late time, it is evolving with negative values i.e. present 
epoch (Fig.1). Where as in case (ii), for $s=0$, from 
equation (\ref{eq45}), we have obtained that, 
at cosmic time $t=\frac{1}{2k_{2}}ln\frac{n^2\alpha^2}{(n-1)L_{0}^2k_{0}^2}$, $\omega=-1$ 
(i.e. cosmological constant dominated universe), when 
$t<\frac{1}{2k_{2}}ln\frac{n^2\alpha^2}{(n-1)L_{0}^2k_{0}^2}$, $\omega>-1$. (i.e. quintessence), 
and when  $t>\frac{1}{2k_{2}}ln\frac{n^2\alpha^2}{(n-1)L_{0}^2k_{0}^2}$, $\omega<-1$. 
(i.e. super quintessence or phantom fluid dominated universe) \cite{ref52}, representing the different phases 
of universe through out the evolving process. Since in both cases, $\frac{\sigma}{\theta}=$ constant, the models 
do not approach isotropy at any time. Therefore, we can not rule out the passibility of anisotropic nature 
of DE at least in Bianchi type III framework. 

\section*{Acknowledgements} 
Authors would like to thank The Institute of Mathematical Science (IMSc), Chennai, India 
for providing facility and support where part of this work was carried out. Author (AKY) is also
thankful to Prof. D. R. Somashekar, Director AEC. Agra, for kind supports.

%\newline
%\nonumsection{References}

\end{document}